\documentstyle[psfig,conf_iap,10pt]{article}
\begin{document}
\heading{%
%Begin Heading
%
The Deimos Spectrograph \\
and a Planned DEEP Redshift Survey \\
on the Keck-II Telescope
%
%End Heading
} 
\par\medskip\noindent
\author{%
%Begin Author names
Marc Davis$^1$
%End Author names
}
\address{%
%First address
Dept. of Astronomy and Physics \\
University of California, Berkeley
}
\author{S. M. Faber$^2$}
\address{Lick Observatory \\
University of California, Santa Cruz}

\begin{abstract}
%Begin Abstract
        A second generation spectrograph for the Keck telescope is under
construction at the Lick Observatory shops and will be delivered
to Hawaii in 1999.  Starting in the Fall of 1999, we shall begin
the second phase of the DEEP project: a dense redshift survey of galaxies
at $z=1$.  With each pointing of 
DEIMOS we shall obtain simultaneous short slit spectra of 70-100 
galaxies with $m_I(AB) < 23.0$ in a field of 15' by 2'.  Four regions of the
sky will be studied in detail, with dense sampling in a region of 120'x15'
in each region, plus outrigger fields.  The galaxies for spectroscopic
analysis will be selected by flux limit and by photometric redshift estimate
$z_{photo}>0.7$.  The goal is to obtain high quality spectra of perhaps 30,000
galaxies over the course of 2-3 years.   I here review the status
of DEIMOS and the science objectives of the survey. 
%End Abstract
\end{abstract}
\section{The Two Faces of DEEP }
%Begin section
The first phase of the DEEP (Deep Extragalactic Evolutionary Probe) survey
has been underway at Keck for the past several years, and has focused on the
properties of faint galaxies.  David Koo earlier this week summarized the
status of that survey, which has been executed with
the first generation Keck spectrograph, LRIS.  The second phase of the
DEEP survey is awaiting completion of the massive DEIMOS (DEEP Imaging
Multi-Object Spectrograph) before it begins in 1999.    Collaborators
on the DEIMOS/DEEP survey with access to the Keck telescope  are 
 David Koo, Garth Illingworth, Tom Broadhurst, Chuck Steidel, and Mark Metzger.
Additional
outside collaborators are Gerard Luppino, Nick Kaiser, Alex Szalay, 
Andy Connolly,
and Richard Kron.  Drew Phillips, Nicole Vogt, and Luc Simard at UCSC, and 
Jeff Newman and Doug Finkbeiner at UCB are also major players in the project.
Sandy Faber is principal investigator for the DEIMOS spectrograph.  Current
details can by found at http://www.ucolick.org/~deep/.

The DEIMOS/DEEP survey will 
be much more extensive than the first generation DEEP survey now underway with
LRIS, 
and will have a number of unique features.  The focus of the survey is to
characterize in detail the nature of the large scale structure of the galaxy
distribution at $z=1$, while at the same time obtaining sufficiently high
quality data to study the internal properties of a large sample of high
redshift galaxies.  Key to both projects is the high spectral resolution that
DEIMOS shall deliver.

\section{The DEIMOS Spectrograph}

The original plan for the DEIMOS spectrograph was to clone four copies of
the LRIS spectrograph together within one package, all using the same
collimator.  This was feasible because the field of view of LRIS is
off-axis from the center of the focal plane of the Keck telescope.  After
a breakthrough in camera design by Harlan Epps, the original four-barreled
DEIMOS evolved into a two-barreled design, with each barrel having twice
the field of view of LRIS, but still sharing a common collimator. Among
other virtues, this brought the fields of view closer to the optical axis
and improved image quality.  For
budgetary reasons, only one barrel is currently under construction, and 
the second barrel, when  eventually built, is likely to be a rather 
different instrument.

DEIMOS is intended for imaging and multislit spectroscopy
over a field of view that is approximately a rectangle of size 15' by 5'.
At the heart of DEIMOS is a monstrous 10-element refractive camera, with
three aspherical surfaces and the largest piece of CaF$_2$ ever cast
(13.5" diameter).  The focal plane of the camera will be paved with silicon
in an array of eight 2k-by-4k CCDs manufactured by the MIT Lincoln Laboratory,
with 15$\mu$ pixels.  The plate scale at the f/15 Nasmyth focus of Keck
is 730$\mu$/arcsec, which is reduced to 126$\mu$/arcsec at the camera focus.
Thus one pixel maps to 0.119 arcsec.  The RMS image diameter 
of the spectrograph 
is anticipated by be $\sim 26\mu$, or 0.21 arcsec.  The format at the focal
plane is square, 8k by 8k pixels, 
and the four CCDs on the red side of the dispersion may 
have thicker substrates with enhanced red response.

LRIS is a Cassegrain instrument in an altitude-azimuth
telescope,  which requires that it 
rotate about the optical axis as the telescope moves across the sky. DEIMOS
is too massive for the Cassegrain focus and 
will reside on a Nasmyth deck of the Keck-II telescope, but it must still
rotate along its long axis.  The time variable gravity loads on DEIMOS
are only in the radial direction, and so the instrument should have
considerably less flexure than LRIS.  Furthermore there is an active 2-D
flexure compensation system within DEIMOS in which a folding mirror and the
X-coordinate of the detector package can be slightly adjusted  by means of
a closed cycle feedback loop. Thus we anticipate that the 
wavelength--to--pixel 
registration within DEIMOS will be stable to  within  0.25 pixel
RMS for long periods of time, hopefully obviating the need to take extensive
calibrations on a daily basis.

DEIMOS is being built as a facility instrument for Keck and will have a
grating slide with room for three diffraction gratings plus a mirror
for imaging.  The slit masks are stored in a juke-box with room for
13 separate masks.  The masks are thin aluminum plates in which slitlets
will be cut by a computer controlled milling machine.  The masks are stored
flat but are bent to conform closely to the curved focal plane of Keck as
they are inserted.

\section{The DEIMOS/DEEP Redshift Survey}
\subsection{Fields and Photometry}
Once DEIMOS is working in 1999, we shall commence a major
redshift survey of faint galaxies designed to characterize galaxies and
the galaxy distribution at a redshift $z=1$.  The intention is to generate
a sample of uniform quality data with a well defined selection criterion that
will be suitable for many different analyses.

\begin{table}
\caption{Fields Selected for the DEEP Survey}
\begin{center}
\begin{tabular}{|c|c|c|} \hline
RA & dec & (epoch 2000) \\ \hline
14$^h$ 17 & +52$^\circ$ 30  & Groth Survey Strip \\ \hline
16$^h$ 52 & +34$^\circ$ 55   & last zone of low extinction \\ \hline
23$^h$ 30 & +0$^\circ$ 00   & on deep SDSS strip \\ \hline
02$^h$ 30 & +0$^\circ$ 00   & on deep SDSS strip \\ \hline
\end{tabular}
\end{center}
\end{table}

This survey will be undertaken
in four fields, as listed in Table 1.  The fields were chosen as low
extinction zones that are continuously observable at favorable zenith 
angle from Hawaii over a six month interval.  One field is the Groth
Survey strip, which has good HST imaging, and two of the fields are on the
equatorial strip that will be deeply surveyed by the Sloan Digital Sky Survey
(SDSS) project.  Each of these fields is the target of a deep imaging survey
by Luppino and Kaiser, whose chief goal is very deep imaging for weak 
lensing studies.  They will use the new UH camera (8k by 12k pixels) with
a field of view of 30' by 40', primarily in the V and I bands, but with
B imaging as well.  The imaging will be obtained in random
pointings spread over
a field of $3^\circ$ by $3^\circ$, but with continuous coverage of a strip
of length $2^\circ$ in the center of each field.

Given this enormous photometric database, we shall use the color information
to make photometric redshift estimates, $z_{photo}$.  DEIMOS will be used
to undertake a spectroscopic survey of galaxies with $m_I(AB) \le 23.0$ and
$z_{photo} > 0.7$.  At this relatively bright flux limit, 2/3 of the
galaxies will have $z<0.7$ \cite{lilly}; the photometric
redshift preselection eliminates this foreground subsample, 
allowing the DEEP project to focus its effort on the high redshift Universe.

\subsection{Choice of Grating and Spectral Resolution}
A choice of gratings is available on DEIMOS, and we anticipate that the
workhorse grating for the DEEP survey will be the 900 lines/mm grating, 
with an anamorphic factor of 1.4.  This grating 
will provide a spectral coverage of 3500 $\AA$ in one setting.  If we use
slits of width 0.75", they will project to a size of 4.6 pixels, or
a wavelength interval of 2 $\AA$.  Thus the resolving power of the observations
will be quite high, $R \equiv \lambda / \Delta\lambda =3700$.

The MIT-LL CCDs have exceptionally low readout noise, 1 $e^-$, and the system
will be sky-noise limited even at high spectral resolution.  The large number
of pixels in the dispersion direction allows us to spread out the
night sky spectrum and to obtain better sky subtraction of the bright OH
sky emission lines.

We will set the grating tilt so that the region 6000-9000 $\AA$ is centered on
the detector, thus assuring that the 3727 $\AA$ [OII] doublet is in range
for galaxies with $0.7 < z < 1.2$.  At the planned spectral resolution, 
the velocity resolution will be 80 km/s, or 40 km/s for an object
with $z=1$.  The [OII] doublet will thus be resolved for all the galaxies,
giving confidence to the redshift determination even if no other features are 
observed.  With sufficient flux it should be possible to measure the velocity 
broadening of the lines, which will hopefully lead to an estimate of the
gravitational potential-well depth of a substantial fraction of the 
galaxies within the survey.

\subsection{Observing Strategy }
%Begin subsection
In each of the four selected fields of Table 1, we shall densely target a
region of 120' by 15' for DEIMOS spectroscopy.  Each pointing of DEIMOS shall
use a unique mask with slitlets cut over a field of size 15' by 2', 
with the slitlets aligned 
along the long axis.  Our goal is 70-100 slitlets per mask, selected from the
list of galaxies with $m_I(AB) \le 23.0$ and $z_{photo} > 0.7$.  The number
density of candidate galaxies will slightly exceed the number of objects we
can select on average, and so our survey will not be 100\% complete.  But
this will not cause problems with the subsequent analysis if we take account
of the positions of those galaxies for which we did not obtain spectroscopy.

The plan is to adjoin 60 contiguous pointings of DEIMOS into a field of
extent 120' by 15', and to supplement these data with 20 additional outrigger
pointings of DEIMOS within the surrouding $3^\circ \times
 3^\circ$ field for  which
photometry is available.   The planned integration time is one hour per pointing,
broken into several shorter integrations for cosmic ray removal.  Because
the slitlets will be very short, we do not plan to dither the telescope. The
stability of DEIMOS should allow excellent sky subtraction in spite of the
short slits.  

The goal of the observing program is to obtain DEIMOS spectroscopy in
$\approx 320$ separate fields, which should net 25,000-30,000 galaxies with
$0.7 < z < 1.2$.  This will require approximately 50 clear nights on Keck,
which we plan to complete over a 2-3 year period.  The data rate while on the
telescope will be approximately 250 Mbytes/hour.

\subsection{Science Goals}
Each of the four DEEP surveys will enclose a densely sampled 
comoving volume of approximately
$500 \times 60 \times 8 h^{-3}$ Mpc$^3$ (in an Einstein--de Sitter cosmology).
The densely sampled portions of the survey will thus be approximately two-dimensional,
but the shortest dimension does exceed the correlation length of the galaxy
clustering.  The outrigger fields will yield information in 
a dilutely sampled cone that is well suited for detection of large scale
filaments.  We have a number of science goals in mind for this database:

\begin{enumerate}
\item
Characterize the linewidths and spectral properties of galaxies versus color,
luminosity, redshift, and other observables.

\item 
Precisely measure the two--point and three--point correlation functions of 
galaxies at $z=1$ as a function of other observables, 
such as color, luminosity,
or linewidth.  For the highe--order correlations,  dense sampling is
essential.  Recent observations of Lyman limit galaxies at $z=3$ \cite{steidel}
suggest that the bias in the galaxy distribution was considerably higher in
the past.  Higher order correlations in the galaxy distribution are one way
to estimate the presence of bias in the galaxy distribution \cite{fry}. If the
galaxy bias is larger at $z=1$ than at present, the correlation
strength of different subsamples of galaxies should show more systematic
variation than is observed for galaxies at $z=0$.

\item
Measure the coherence of Large Scale Structure at $z=1$, in comparison to the
structure observed at $z=0$ from existing redshift surveys.  Different 
cosmological models make very different predictions for the appearance of
the high redshift LSS \cite{frenk}.  
Differing degrees of bias in the galaxy distribution 
will be apparent in such images.

\item
Measure redshift space distortions in the galaxy clustering at $z=1$ by
means of the $\xi(r_p,\pi)$ diagram and by direct measure of the small
scale thermal motions of galaxies, such as suggested by \cite{davis}.  The
evolution of the thermal velocity dispersion is another handle that can
separate the evolution of the galaxy bias from the evolution of the 
underlying matter distribution.  The high redshift precision expected from
DEIMOS will make this measurement possible.

\item
Separate the Alcock-Paczynski effect \cite{alcock},\cite{ball} from 
the redshift 
space distortions of the $\xi(r_p,\pi)$ diagram.  This effect relates
intervals of angular separation versus intervals of redshift separation 
as a function of redshift. 
An object that appears spherical at low redshift would
appear elongated in redshift at $z>0$, but the degree of elongation is
a function of $q_0$.  It is just conceivable that the DEEP project will
provide data that can measure this effect and separate it from the
other expected redshift space distortions.

\end{enumerate} 

\section{The Next Three Years} 

In summary, the completion of DEIMOS will signal the initiation of a massive
survey of faint galaxies with the Keck telescope.  This next phase of DEEP
will not push to the ultimate flux limit of Keck, but will use the enormous
light gathering power of the telescope combined with the unprecedented
field of view to undertake a detailed examination of the state of galaxies
and galaxy clustering at $z=1$.  The proposed studies require dense sampling
and precision redshifts of the galaxy distribution.  With these data,
it should be possible to break the degeneracy between $b(z)$, $\sigma_8(z)$,
and $\Omega(z)$.  With a bit of luck, we may also obtain a separate
measure of the deceleration parameter, $q_0$.  The next few years promise
to be extremely busy, but very exciting.

% Begin acknowledgements
\acknowledgements{This work was supported in part by NSF grants AST95-28340 
and AST95-29028. The DEIMOS spectrograph is funded by a grant from CARA
(Keck Observatory), by an NSF Facilities and Infrastructure grant (AST92-2540), by the Center for Particle Astrophysics, and by gifts from Sun Microsystems
and the Quantum Corporation.}
% End acknowledgements

%References should be refered as : \cite{LH}, \cite{MMM}, and \cite{Kea}. 

%
\vfill

\begin{iapbib}{99}{
% Begin bibliography
\bibitem{lilly} Crampton, D., Le Fevre, O., Lilly, S.J., \& Hammer, F. 1995
\apj, 455, 96
\bibitem{steidel} Giavalisco, M., Steidel, C. C, Adelberger, K., Dickinson, M.,
Pettini, M., \& Kellogg, M. 1998, astro-ph/9802318
\bibitem{fry} Fry, J. 1996, \apj Lett., 461, 65L  
\bibitem{frenk} Jenkins, A., Frenk, C. S., Pearce, F. R., Thomas, P. A., 
Colberg, J. M., White, S. D. M., Couchman, H. M. P., Peacock, J. A., Efstathiou,
G., \& Nelson, A. H. 1997, \apj, 499, 20
\bibitem{davis} Davis, M., Miller, A., \& White, S.D.M. 1997, \apj, 490, 63
\bibitem{alcock} Alcock, C., \& Paczynski, B. 1979, Nature, 281, 358
\bibitem{ball} Ballinger, W. E., Peacock, J. A., \& Heavens, A. N. 1979,
 MNRAS, 282, 877
}
\end{iapbib}
\end{document}